# Emergency Powers in Response to COVID-19:
# Policy diffusion, Democracy, and Preparedness

Magnus Lundgren
Stockholm University
magnus.lundgren@statsvet.su.se

Mark Klamberg
Stockholm University
mark.klamberg@juridicum.su.se

Karin Sundström
Stockholm University
karin.sundstrom@statsvet.su.se

Julia Dahlqvist
Stockholm University
julia.dahlqvist@juridicum.su.se

# I. Introduction

The COVID-19 pandemic has prompted different responses ranging from what could be described as "elimination", "suppression", "mitigation", to "inaction". From an international law perspective, the pandemic raises questions what state actions are required, permitted and prohibited. Depending on how they are designed, states' responses to combat COVID-19 may affect various human rights, e.g. the right to life, the right to health, freedom of movement, and freedom of assembly.

One option available to states is to declare a state of emergency (SOE). During times of crisis, derogating from their human rights treaty obligations and declaring a SOE may assist some states to effectively combat the situation. However, while emergency powers can be necessary and legitimate in certain cases, they can also be abused. When public health is seriously imperiled, human rights law is already permissive, allowing for a wide range of measures without declaring a SOE.

In response to the COVID-19 pandemic, states made different choices: About half declared a SOE during the first half of 2020 while the other half did not. Why did states facing the same threat follow such different strategies? The article suggests and evaluates three potential explanations to this puzzle: First, states look to their regional peers for inspiration and legitimation, leading to patterns of regional policy diffusion. Second, newer and less robust democracies are more likely to resort to SOEs, compared with mature democracies and dictatorships. Third, states with a higher pandemic preparedness are less likely to resort to a SOE.

This paper begins with giving an account of the applicable legal frameworks, including the World Health Organization (WHO) International Health Regulations (IHR 2005) and the International Covenant on Civil and Political Rights (ICCPR[1]). It then presents theoretical arguments for why states may be motivated to declare a SOE in response to the COVID-19 pandemic. This provides the basis for the subsequent data analysis. Based on a statistical analysis of panel data covering 180 countries, the paper finds evidence that regional policy diffusion and regime characteristics may explain why states declare SOE while preparedness offers less of an explanation.

## II. International Law and Pandemics

When examining the use of SOE as a response to the COVID-19 pandemic, different legal frameworks interact, and concrete rules may serve to protect a certain value, enable, require or restrict, states in their actions. The right to health, like all human rights, imposes three obligations on States parties: the obligations to *respect*, *protect*, and *fulfil*. The obligation to *respect* requires, inter alia, States to refrain from using or testing biological weapons. The obligation to *protect* requires, for example, States to take measures that prevent third parties limiting people's access to health-related information and services. Finally, the obligation to *fulfil* requires States to adopt measures to promote the realization of the right to health.[2] The obligation to *fulfil*,

---

[1] International Covenant on Civil and Political Rights, adopted 16 December 1966, 999 UNTS 171.
[2] CESCR General Comment No. 14: The Right to the Highest Attainable Standard of Health (Art. 12)

which is the most relevant here, includes preparing the health care system for a pandemic, to act during a pandemic, and, if necessary, to declare a SOE. The IHR 2005 and rules on SOEs could be described as operationalizations of the State obligation to fulfil the right to health.

*WHO 2005 Regulations*

Global pandemics trigger questions under global health law. Traditionally, crisis management and public health powers reside in sovereign states where international law serves as protection against abuse by states. The prompt identification and control of diseases requires both national and international action. The WHO is at the center of the global health regime.[3]

The World Health Assembly may pursuant to Article 21 of the WHO constitution adopt regulations concerning "sanitary and quarantine requirements and other procedures designed to prevent the international spread of disease". In 2005 the WHO adopted IHR 2005 which places emergency authority — to the extent states delegate such to the WHO — in the hands of the Secretariat. The IHR 2005 is one of the world's most widely adopted treaties, requiring its 196 parties to build capacity to detect, assess, notify, and respond.[4] Pursuant to IHR 2005 Article 6(1), each State

---

Twenty-second Session of the Committee on Economic, Social and Cultural Rights, 11 August 2000, E/C.12/2000/4, paras. 33-36.
[3] J. Benton Heath, *Global Emergency Power in the Age of Ebola*, 57 HARV. INT'L L.J. 1, 1-2, 6-7 (2016).
[4] Lawrence O. Gostin, Global Health Law (Harvard University Press, 2014), at 182.

Party must monitor health hazards occurring within their territories and notify WHO of all events which may constitute a public health emergency of international concern (PHEIC). The WHO Director-General, pursuant to regulation 12, has exclusive power to determine and declare whether an event constitutes a PHEIC.

The criteria for declaring a PHEIC are difficult to frame in purely legal terms.[5] The system has been criticized for its binary approach, with the declaration of a PHEIC as the only level of alert, and for relying on an opaque assessment procedure.[6] WHO declared COVID-19 to be a PHEIC on 30 January 2020.

*Restrictions of human rights during normal peace-time conditions*

Human rights law protects several different rights, not all of which are absolute. The functions of society and human interaction arguably require that some human rights may be restricted under certain circumstances. Rights such as freedom of movement, freedom of assembly and right to property have clauses which allow interference with the concerned human right under normal peace-time conditions for public reasons. To restrict such rights, the interference must: have a legal basis (a law normally adopted by parliament), pursue a legitimate aim, be necessary in a democratic society, and be proportionate.[7] This means that when a state is faced with a pandemic, such as COVID-19, substantial restrictions of rights are possible without declaring a SOE. The utility of a declaration of SOE is that measures may be adopted in a more expeditious way by Government. In this regard, Scheinin has argued that it is "the safe course of action to insist on the principle of normalcy, i.e. to handle the crisis through normally applicable powers and procedures and insist on full compliance with human rights."[8]

---

[5] Pedro A. Villarreal, 'Public International Law and the 2018-2019 Ebola Outbreak in the Democratic Republic of Congo' *EJIL: Talk*, 1 August 2019; Burci, Gian Luca, 'The Outbreak of COVID-19 Coronavirus: are the International Health Regulations fit for purpose?' *EJIL: Talk*, 27 February 2020.
[6] Burci, *supra* note 5.
[7] See ICCPR Art. 17-19, 21-22.

*State of Emergency*

The rationale for a SOE is to increase a state's ability to safeguard the life of the nation in exceptional situations. The declaration of a SOE may involve modifying ordinary laws, special emergency legislation, or interpretative accommodation (by judges) of the constitution and laws.[9]

There are two main approaches regarding whether such a suspension can be contained within the judicial order. The *rule-or-law* approach holds that measures to counter a crisis must be within the confines of law, while the *sovereignty approach* argues that emergency measures by their nature cannot be reduced to legal norms.[10]

---

[8] Martin Scheinin, 'To Derogate or Not to Derogate?' *Opinio Juris*, 6 April 2020.
[9] OREN GROSS & FIONNUALA NÍ AOLÁIN, LAW IN TIMES OF CRISIS: EMERGENCY POWERS IN THEORY AND PRACTICE 2 (CAMBRIDGE UNIVERSITY PRESS 2006), at 66-79.
[10] Giorgio Agamben, STATE OF EXCEPTION (TRANSLATED BY KEVIN ATTELL (The University of Chicago Press 2005), at 10, 22-23; Scott P. Sheeran, *Reconceptualizing States of Emergency under International Human Rights Law: Theory, Legal Doctrine, and Politics*, 34 MICH. J. INT'L L. 491, at 500.

SOEs may be regulated at different levels: the international, national, and sub-national.

With the inclusion of derogation clauses, international human rights law allows SOEs under certain conditions. The purpose is to provide authorities extraordinary powers and resources, and to liberate the sovereign from legal constraints. However, human rights law serves to provide checks on sovereign powers to protect individuals, especially in times of crisis.

The prerequisites for derogation from obligations under the ICCPR are provided by Article 4. The article in its first paragraph gives that (1) the invoked public emergency must threaten the life of the nation, (2) a SOE must be officially proclaimed, (3) the measures should be limited to what is strictly required, (4) the measures should be consistent with other obligations under international law, and (5) the measures cannot involve discrimination solely based on e.g. race and religion. The second paragraph gives that derogation from certain rights is prohibited. Furthermore, states who invoke Article 4 immediately have to notify other State Parties, through the Secretary-General of the United Nations, which provisions it has derogated from, and the reasons for the decision.

Not all public emergencies are ground for derogation.[11] In relation to COVID-19, the question arises whether the pandemic could "threaten the life of the nation", and thereby constitute ground for derogation. This distinction needs to be made in relation to each individual country. However, guided by previous definitions it is likely that COVID-19, as it has resulted in widespread illness and deaths affecting a significant part of the population, could be a threat within the scope of Article 4.[12] In April 2020, the United Nations Human Rights Office issued a guidance, stating inter alia that "[i]f derogations from a State's human rights obligations are needed to prevent the spread of COVID-19 […]", indicating that the pandemic *could* constitute a threat to the life of the nation.[13]

---

[11] United Nations Human Rights Committee, CCPR General Comment No. 29: Article 4: Derogations during a State of Emergency, 31 August 2001, CCPR/C/21/Rev.1/Add.11, para. 3.

On the national level, several constitutions permit and differentiate between types of emergencies. For example, the constitutions of the Netherlands and Portugal establish a dual structure where the former distinguishes between "state of war" and a "state of emergency," and the latter between a "state of siege" and a "state of emergency." SOE may also be regulated at a sub-national level, with the United States being one example.[14] In contrast to the federal constitution, many of the state constitutions contain more explicit emergency provisions.[15]

---

[12] See e.g. The Siracusa Principles on the Limitation and Derogation Provisions in the International Covenant on Civil and Political Rights, 28 September 1984, E/CN.4/1985/4, II(A) principle 39.
[13] OHCHR, Emergency Measures and Covid-19: Guidance, 27 April 2020, at https://www.ohchr.org/Documents/Events/EmergencyMeasures_COVID19.pdf.
[14] Henry P. Monaghan, *The Protective Power of the Presidency*, 93 COLUMBIA LAW REVIEW 1993, 1-74, at 32-38; George Winterton, *The Concept of Extra-Constitutional Executive Power in Domestic Affairs*, 7 HASTINGS CONST. L.Q. 1, 24-25 (1979); GROSS & NÍ AOLÁIN, *supra* note 6, at 37-38.
[15] Oren Gross, *Providing for the Unexpected: Constitutional Emergency Provisions*, 33 ISRAEL YEARBOOK ON HUMAN RIGHTS 13, 19-20 footnote 28 (2003).

**III. Explaining states of emergency**

In handling pandemics or other emergencies, states have a repertoire of policy choices. Out of this menu, some states will be more likely than others to choose to declare a SOE in the face of a pandemic like COVID-19. This propensity, we suggest, is based on a calculation of the costs and benefits of such a declaration. Our general framework draws on previous research on SOEs and extends it to develop expectations regarding how states employ SOEs during pandemics.[16]

Quantitative research has examined the circumstances that make SOEs more or less likely. Bjørnskov and Voigt find that SOEs have different sources depending on if they take place in the context of natural disasters or political turmoil.[17] Constitutions matter as states lacking emergency powers declare SOEs more often than those that have them. When it comes to natural disasters, research suggests that executives facing low costs from SOEs are more likely to declare them. If a country is undergoing an economic crisis, an SOE is more likely following political turmoil. An SOE is also more likely after natural disasters when the legislature holds more power.[18] A study on countries with "Western-style" constitutions finds that terrorist incidents are associated with a higher likelihood of a SOE, but that they are less likely in election years.[19] Next to the quantitative research, case studies have found, for example, that SOEs can evolve from temporary measures, as was the case in Egypt, to become the routine mode of governance, and that they, far from being unpopular, can be legitimized by emotions of compassion, as seen in Venezuela following a devastating landslide.[20]

---

[16] Christian Bjørnskov and Stefan Voigt, *Why Do Governments Call a State of Emergency? On the Determinants of Using Emergency Constitutions*, 54 EUROPEAN JOURNAL OF POLITICAL ECONOMY 110 (2018).
[17] *Ibid.*
[18] *Ibid.*

We assume that states did not have an interest in declaring a SOE before the pandemic (unless they were already under a SOE), but that their calculations may shift in response to COVID-19. This shift, we suggest, arises primarily as states balance their responsibility to respect, protect, and fulfil the rights of their citizens. Under some circumstances, a SOE may be seen as a necessary tool. A SOE may offer legitimacy for the state to take actions otherwise impossible (or with a considerably higher price), or the ability to undertake necessary measures, e.g., lockdown, due to a lack of other alternatives. The granting of legitimacy and support, we suggest, may be particularly pronounced during a pandemic in the face of human suffering. Even during exogenous events like a pandemic, however, an SOE comes with costs. These include unpopularity, especially if seen as an exaggerated response, and challenges in declaring a SOE, for example difficulties in securing support in parliament.

---

[19] Christian Bjørnskov and Stefan Voigt, *When Does Terror Induce a State of Emergency? And What Are the Effects?*, 64 JOURNAL OF CONFLICT RESOLUTION 579 (2020).
[20] Sadiq Reza, *Endless Emergency: The Case of Egypt*, 10 NEW CRIMINAL LAW REVIEW 532 (2007). Didier Fassin and Paula Vasquez, *Humanitarian Exception as the Rule: The Political Theology of the 1999 Tragedia in Venezuela*, 32 AMERICAN ETHNOLOGIST 389 (2005).

We expect that a government's weighing of costs and benefits from a SOE are shaped by three key factors: the regional acceptance of SOEs, the robustness of its democratic institutions, and its pandemic preparedness.

States are often influenced by each other, through emulation and learning, which can lead to processes of diffusion and patterns of convergence as states gravitate towards the same policy solutions.[21] Following this logic, we expect that a state's deliberation about whether to declare a SOE is shaped by the decisions of other states, in particular by those that are geographically proximate. In a typical diffusion pattern, we would expect to see an S-shaped curve, with few early declarations, many declarations in close proximity, and few laggards.[22] This logic motivates the following hypothesis:

> Hypothesis 1: *The more states within a region that have declared SOEs, the more likely states in the same region are to declare SOEs.*

Viewing SOEs as an action that may restrict the rights and liberties of citizens leads us to expect that some states will be particularly prone to use this policy.[23] Previous research suggests that we may expect "more in the middle" when it comes to the relationship between the strength of democratic institutions and repression.[24] Accordingly, SOEs may be less likely in established democracies and strong autocracies, and most likely in states in between these two extremes. Autocracies may enjoy SOE-equivalent powers without having to declare an actual SOE, and democracies are likely to find SOEs, anathema to central principles of liberal governance and unappealing to electorates, as too costly. This leaves us with the states "in the middle", so called anocracies, which face less restraints from electorates than democracies but also lack access to the kind of repressive techniques available to autocracies. This gives us the following hypothesis:

> Hypothesis 2: *States with weak democratic institutions are more likely to declare a SOE compared to democracies or autocracies*

---

[21] See e.g. Katharina Holzinger and Christoph Knill, *Causes and Conditions of Cross-National Policy Convergence*, 12 JOURNAL OF EUROPEAN PUBLIC POLICY 775 (2005).
[22] Virginia Gray, *Innovation in the States: A Diffusion Study*, 67 AMERICAN POLITICAL SCIENCE REVIEW 1174 (1973).
[23] See e.g. Christian Davenport, *State Repression and Political Order*, 10 ANNUAL REVIEW OF POLITICAL SCIENCE 1 (2007).

Finally, because state preparedness provides a more resilient health infrastructure, it may obviate the need to declare a SOE. Higher preparedness entails better ability to handle the pandemic before it becomes necessary to respond with an SOE and offers access to alternative means, such as track-and-trace systems, which may serve as a substitute for a SOE. Tying back to the previous discussion, this suggests that states that have come further in respecting and fulfilling the right to health will be less likely to declare a SOE. This gives us the following hypothesis:

> Hypothesis 3: *States with higher pandemic preparedness are less likely to declare a SOE*

---

[24] Helen Fein, *Life-Integrity Violations and Democracy in the World, 1987*, 17 HUMAN RIGHTS QUARTERLY 170 (1995).

## IV. Data and methods

To attain data on COVID-19-related SOEs, we sourced information from two public datasets and carried out our own supplementary research. The Centre for Civil and Political Rights (CCPR) provides data on 136 parties to the ICCPR.[25] A "Covid-19 Civic Freedom Tracker" maintained by the International Center for Not-for-profit Law (ICNL) identifies 87 SOEs declared in response to COVID-19, some of which are not reported in the CCPR dataset.[26] To address missingness in these two sources, we collected data on 54 states using news and online sources.[27] The combined dataset covers 180 states, providing information on whether or not they declared a SOE between January 1, 2020, to June 12, 2020.

    Figure 1 illustrates the geographic scope of these data. We observe that 101 declared SOEs and 79 did not, while a smaller group either had pre-existing SOEs

---

[25] COVID-19 State of Emergency Data (visited June 28, 2020), at https://datastudio.google.com/reporting/1sHT8quopdfavCvSDk7t-zvqKIS0Ljiu0/page/dHMKB.
[26] COVID-19 Civic Freedom Tracker (visited June 28, 2020) at https://www.icnl.org/covid19tracker/.
[27] We are grateful to Martin Lundqvist for research assistance.

(e.g. Egypt) or could not be determined. We also note signs of spatial clustering, with some regions (e.g. Latin America) displaying widespread emergencies, whereas others (e.g., South Asia) containing no or very few emergencies.

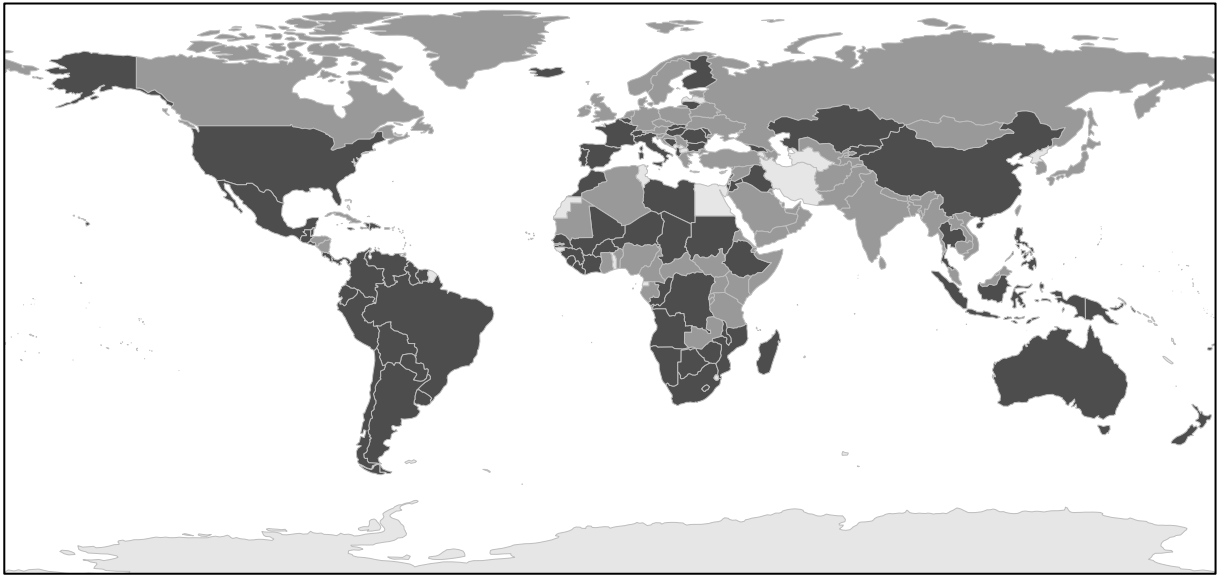

**Figure 1**. Geographic distribution of COVID-19-related SOEs. Data: CCPR (2020), ICNL (2020), and authors' own. States with declared SOEs marked in dark grey. States without SOEs marked in grey. States where information is missing or where SOEs existed prior to COVID-19 marked in light grey.

Figure 2 exhibits the proportion of states with SOEs by day since January 1, 2020. We note three key patterns. First, SOEs were rare in the first three months after the virus became publicly known, including after WHO declared it a PHEIC. The first country to declare a SOE was Italy, one of the earliest and hardest hit countries, which did so on January 31. Second, following WHO's declaration of a pandemic on March 11 (but not necessarily because of it), there was a rapid growth in SOEs. In fact, nearly all SOEs, regardless of region, were declared during a month-long period from early March to early April. Third, there were few new SOEs in the following two months.

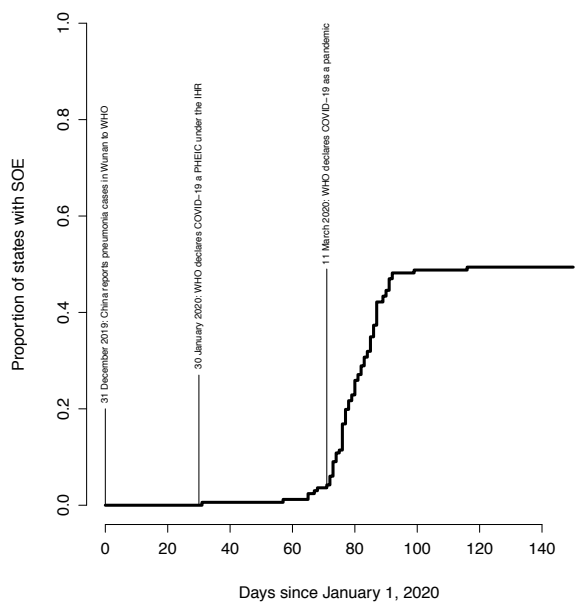

**Figure 2**. Proportion of states with declared SOEs, by day since January 1, 2020.

Taken together, the patterns observed in Figures 1 and 2 suggest that SOEs exhibit clustering in space and time. The spatial pattern is consistent with a geographic diffusion logic in which states are influenced by their regional peers. The temporal clustering suggests that states went through a period where they shared a simultaneous sense of urgency, leading them to similar policy responses despite having had widely different exposure to the pandemic thus far.

While the rapid spread and spatial homogeneity are consistent with a diffusion logic, the fact that many states never adopted SOEs suggests that any such process was conditioned on other factors. To account for such factors and to identify the determinants of SOEs, we use multiple regression, expanding our dataset to include variables for our privileged theoretical expectations of regional diffusion, democracy, and pandemic preparedness, as well as a range of control variables.

To represent our diffusion logic, we create the variable *regional SOE*, operationalized as the cumulative count of SOEs in a given geographic region. We follow World Bank definitions and identify seven main regions (see Figure 4). States in regions where SOEs are prevalent will score high on this variable and vice versa.

We operationalize *democracy* based on the V-Dem liberal democracy index.[28] The index judges the quality of democracy by the limits placed on government and the extent to which it protects the rights of individuals and minorities. The index factors in the protection of civil liberties, rule of law, the independence of the judiciary, and limitations on the executive. States with stronger democratic institutions have higher values on this variable.

---

[28] Michael Coppedge, et al. 2019. *V-Dem Dataset v9*. Varieties of Democracy (V-Dem) Project.

We measure *pandemic preparedness* based on the global health security index, which summarizes information on states' relative capability to prevent and mitigate pandemics.[29] States with strong pandemic preparedness score high on this index and vice versa.

We control for key background factors. To reflect countries' varying institutional contexts, we control for democratic durability (years since democratic transition), historical SOEs, and GDP per capita. To account for immediate anti-COVID-19 measures, we adjust for the stringency of governmental actions using data from the Oxford COVID-19 government response tracker.[30] To account for demographic variation that may shape how gravely a government views the pandemic threat, we control for population above 65 years of age, life expectancy, and population density. Finally, to account for variation in the pandemic pressure, we control for cumulative national deaths from COVID-19.[31]

We construct a time-series cross-sectional data structure with daily observations on each state between January 1 and June 12, 2020. The dependent variable (SOE or not), regional SOEs, policy stringency, and cumulative deaths vary across both days and states; other variables vary across states. We use a logit estimator, capture temporal effects with a cubic polynomial, and add country random effects to account for unobserved heterogeneity. For presentational purposes, we estimate some models using a cross-sectional structure, with observations on each state on June 12, 2020.

---

[29] Global Health Security Index (visited June 29), at https://www.ghsindex.org.
[30] Thomas Hale, et al. 2020. Variation in government responses to COVID-19. BSG-WP-2020/32. May 2020.
[31] Using alternative measures of pandemic pressure, such as the total number of infected or hospitalized, do not affect the results.

**V. Results**

Our key results are summarized in Figure 3.[32] The three panels illustrate the predicted probability of SOEs as a function of our three key variables, holding all other factors constant. In other words, for each value of an explanatory variable (plotted on the x-axes), the figures show the probability that a state with such characteristics will have adopted a SOE during the period of study.[33]

---

[32] Full results and regression tables are reported in the online appendix.
[33] For ease of interpretation, we show results based on a cross-sectional model; equivalent results are attained in the time-series cross-sectional model.

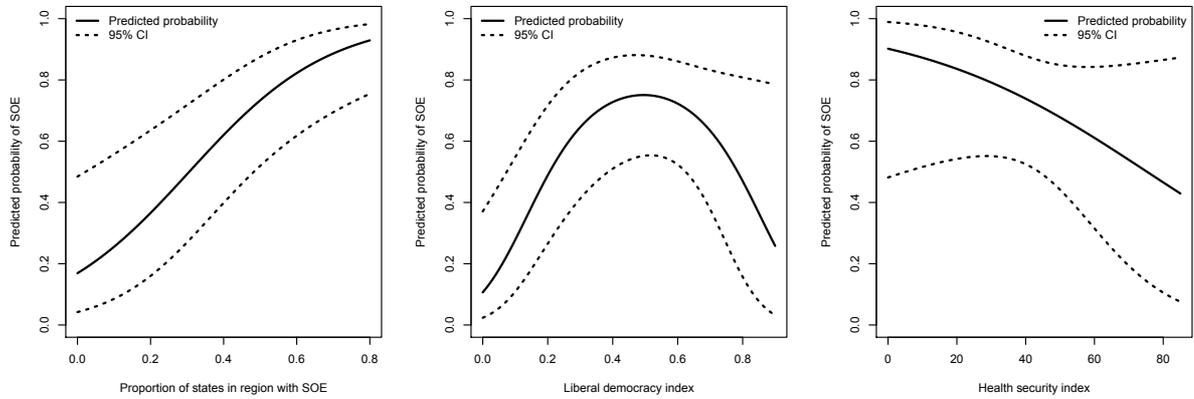

**Figure 3**. Predicted probabilities of SOEs, as a function of (a) regional SOEs; (b) democracy; and (c) pandemic preparedness. Dashed lines represent 95 percent confidence intervals.[34]

Our first hypothesis was that of a regional diffusion logic: When contemplating the menu of possible policy responses to the onslaught of COVID-19, states would look to their regional peers for inspiration. Our data bear this expectation out. Our regression models indicate that, among countries with comparable characteristics and disease spread, those having a high number of regional peers with declared SOEs are significantly more likely to declare a SOE themselves, and vice versa.

---

[34] Calculations based on a logit model with random country effects. All continuous variables held constant at their means and categorical variables set at their reference values.

Figure 4 illustrates these dynamics in greater detail. We plot the declaration of SOEs over time and disease spread. Reinforcing the patterns discussed in relation to Figures 1 and 2, we observe that regions exhibit a tendency towards homogeneity in SOE declarations. This is particularly clear for Latin America (especially South America) and South Asia (lacking SOEs). It is also clear that these regional diffusion processes happen around the same time, calendar-wise, but at very different points in time on the pandemic curve. In East Asia, SOEs were declared after the pandemic curve has already started to flatten out, whereas in Latin America, Sub-Saharan Africa, and the Middle East, SOEs were declared at considerably earlier points of the curve, signifying lower levels of pandemic impact, and in many cases before there had been any local impact at all.[35]

---

[35] In 50 states, SOEs were declared before there was a single local death from COVID-19.

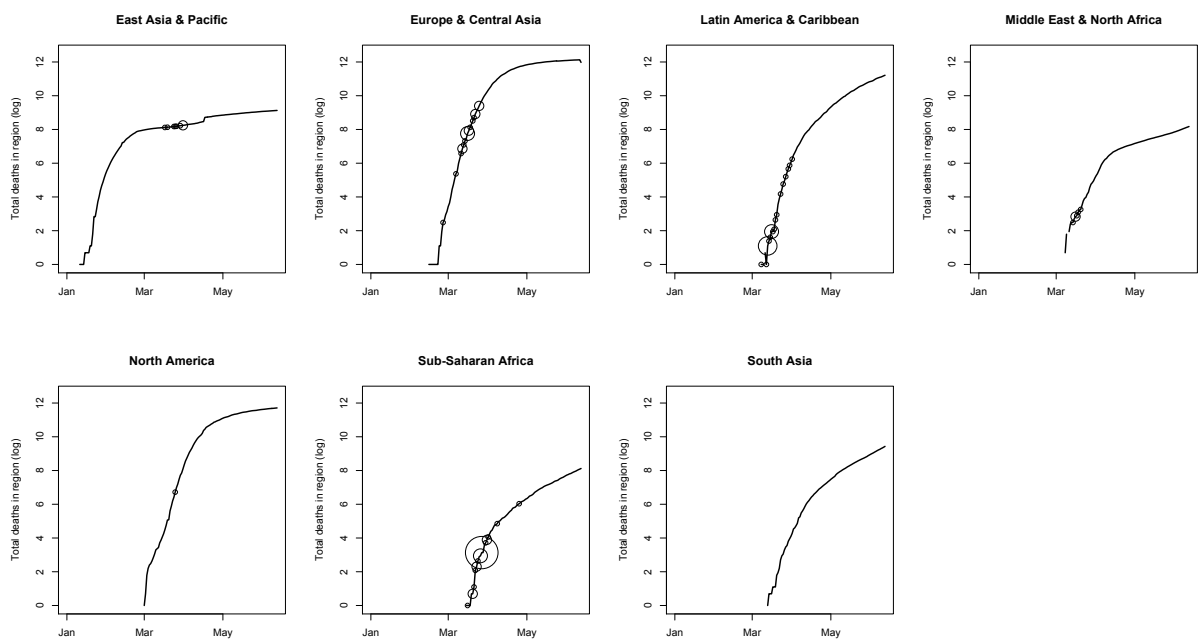

**Figure 4**. Declaration of SOEs and COVID-19 deaths across geographic regions, January-June 2020. Size of rings is proportional to the number of states in a region declaring a SOE on a given date.

Our second expectation was that states' willingness to declare SOEs would vary with their institutional characteristics: In robust democracies, the costs of SOEs would outweigh the benefits; in autocracies, where governments already enjoy many of the powers that SOEs can enable in liberal societies, a SOE would be redundant. Only in the "middle," characterized by newer and less robust democracies, would the benefits of SOEs outweigh the costs. As is evidenced by the inverted U-shape pattern in panel

(b) of Figure 3, this expectation is born out in the data. Indeed, it is the strongest and most robust of our findings.[36]

In our third hypothesis, we proposed that states with a higher preparedness would be more able to weather COVID-19 without resorting to emergency powers, whereas under-prepared countries would see SOEs as a necessary tool to combat the spread of the disease. This hypothesis receives only weak support in the data. There is indeed a negative correlation between preparedness and the likelihood of SOEs, in line with theory, but it is not statistically significant at the 95 percent level. Relatedly, we find that states that have already adopted more stringent policies are also more likely to adopt a SOE.

Overall, we view the results as favoring hypothesis 1 and 2 while hypothesis 3 remain uncertain. These results are robust to a number of alternative estimation strategies, including to controlling for constitutional emergency powers, using alternative measures of democracy (Polity IV data), defining regions differently (by continent), and excluding states with federal systems or within-state variation in SOEs (e.g., the United States).

---

[36] Trinidad and Tobago (TTO) illustrates how comparatively robust democratic institutions may affect the decision to declare an SOE. Based on our statistical models, TTO was very likely to opt for an SOE: The country is situated in Latin America, suggesting a permissive environment, and its preparedness is below average. Yet it did not. A likely explanation is TTO's democratic institutions, which are stronger than most of its regional peers (its liberal democracy index is 0.63, compared with 0.49 for the region). Qualitative evidence aligns with this interpretation. For example, the Prime Minister of TTO considered that a SOE was not necessary to handle COVID-19 and "would only give additional powers to police and a reduction of citizens' rights and privileges," while an opposition leader declared that the "wide reaching detention powers" of a SOE "should only be used to restore law and order".

Our control variables suggest several interesting results. Most importantly, we find no significant association between pandemic impact, measured as national COVID-19-related deaths, and SOEs. Aligning with the descriptive finding that many states proclaimed SOEs before the disease spread locally, this suggests that many states employed emergency powers proactively, seeking to bolster preparedness before the storm of the pandemic arrived. The clearest exception to this pattern is France, which suffered nearly 100 dead to COVID-19 before it declared a SOE on March 22, 2020.

**VI. Conclusions**

Using a combination of existing and novel data, we examined empirical patterns in COVID-19-related SOEs, seeking to explain why some states resorted to emergency powers while others did not. The results suggest that states' declaration of SOEs is driven by both external and internal factors. A permissive regional environment, characterized by many and simultaneously declared SOEs, may have diminished reputational and political costs, making employment of emergency powers more palatable for a wider range of governments. At the same time, internal characteristics, specifically democratic institutions and pandemic preparedness, shaped governments' decisions. Weak democracies with poor pandemic preparedness were considerably more likely to opt for SOEs than dictatorships and robust democracies with higher preparedness. We find no significant association between pandemic impact, measured as national COVID-19-related deaths, and SOEs, suggesting that many states adopted SOEs proactively before the disease spread locally.

These results have several implications for research on international law and SOEs. First, our data indicate that only a small minority of states has notified the United Nations Secretary General despite ICCPR requirements. This raises important questions for the research on the effectiveness of international law. Second, the finding that states with weaker democratic institutions, in between autocracies and democracies, have been more likely to declare SOEs speaks to classical debates in the study of democratization and the rule of law. Most notably, it raises questions for the study on whether constitutions can amount to credible commitment to the protection of liberal rights or whether they serve as legitimization devices without real bite. Finally, the finding that robust democracies and autocracies have been less willing to invoke a SOE points to possible variation in the ability of states to handle the pandemic. Although this article has not examined the question of what is the most sensible state response to COVID-19, it is noteworthy that international human rights law allows very strict measures as a response to the COVID-19 pandemic, including a lockdown and stay-at-home orders – without declaring a SOE.